\documentclass[aip, sd, amsmath, amssymb, reprint]{revtex4-1}

 \usepackage[english]{babel}
 \usepackage{graphicx}
 \usepackage{dcolumn}
 \usepackage{bm}
 \usepackage{color} 
 \usepackage[caption=false]{subfig}

\begin{document}
	
\title{Colloidal Motion under the Action of a Thermophoretic Force}

\author{Jerome Burelbach}
\email{jb920@cam.ac.uk, ee247@cam.ac.uk}
\affiliation{Cavendish Laboratory, University of Cambridge, Cambridge CB3 0HE, United Kingdom}
\author{Mykolas Zupkauskas}
\affiliation{Cavendish Laboratory, University of Cambridge, Cambridge CB3 0HE, United Kingdom}
\author{Robin Lamboll}
\affiliation{Cavendish Laboratory, University of Cambridge, Cambridge CB3 0HE, United Kingdom}
\author{Yang Lan}
\affiliation{Cavendish Laboratory, University of Cambridge, Cambridge CB3 0HE, United Kingdom}
\author{Erika Eiser}
\email{jb920@cam.ac.uk, ee247@cam.ac.uk}
\affiliation{Cavendish Laboratory, University of Cambridge, Cambridge CB3 0HE, United Kingdom}

\begin{abstract}

We present thermophoretic measurements in aqueous suspensions of three different polystyrene
(PS) particles of varying negative charge, size and surface coating. Our measurement technique is based on the observation of the colloidal steady-state distribution using conventional bright-field microscopy, which avoids undesirable effects such as laser-induced convection or local heating. We find that the colloids with the weakest zeta potential exhibit the strongest thermophoretic effect, suggesting that surface functionality leads to a more intricate dependence of the Soret coefficient on hydrodynamic boundary conditions than predicted by existing theoretical approaches. We also study the relaxation of the colloids to steady-state and propose a model to quantify the relaxation speed, based on the time evolution of the colloidal center of mass. Our observations are well described by this model and show that the relaxation speed tends to increase with the magnitude of the thermophoretic force.
\end{abstract}

\maketitle

\section{Introduction}

The motion of particles in a temperature gradient is known as thermophoresis.
The thermophoretic effect has been studied in a wide range of systems, from charged particles \cite{Putnam2005,Duhr2006b,Iacopini2006,Braibanti2008,Duhr2006,Dhont2007} or micelles \cite{Piazza2002,Piazza2003}
in aqueous electrolyte solutions to polymers in polar and
nonpolar solvents \cite{Schimpf1987,Zhang1999,Duhr2004,Braun2002}. Experimental evidence suggests that
the thermophoretic velocity is insensitive to particle size, making
it the ideal candidate for the fractionation of small biomolecules
\cite{Jeon1997}, as opposed to dielectrophoresis or magnetophoresis, where
the velocity scales with the square of the particle radius \cite{Piazza2008}.
It has also been shown that thermophoresis combined with convection
can be used as a focussing technique to achieve strong accumulation
of DNA \cite{Braun2002}, indicating that it might have played a fundamental
role in the formation of life \cite{Baaske2007}. 

Despite these advances, studying thermophoresis in colloidal suspensions
remains a challenging task, both experimentally and theoretically. Thermophoresis
in dilute suspensions is driven by hydrodynamic stresses resulting
from a local interaction between colloid and fluid. This interaction is influenced by a wide range of parameters \cite{Piazza2008},
including intensive variables of the system such as temperature, salinity,
pH and solvent expansivity, as well as single colloid properties like
shape and surface coating \cite{Wurger2010}. Some experimental techniques suffer from undesirable effects that inhibit
the direct measurement of purely thermophoretic motion, such as local
heating or convection induced by external fields. Here, we report measurements based on the observation of the colloidal steady-state distribution in a closed cell, using conventional bright-field microscopy. This method is practically bias-free and has the advantage of capturing all single-particle and collective contributions to thermophoresis. Furthermore, we study the motion of colloids during the relaxation to steady-state and propose a theoretical model to describe this relaxation more quantitatively.

The motion of colloids resulting from thermodynamic gradients in concentration and temperature is quantified by the total particle
flux $\vec{J}$, which is given by \cite{Piazza2008}:
\begin{equation}
\vec{J}=-D\nabla c-cD_{T}\nabla T,\label{eq:-1}
\end{equation}

where $D$ is the Fickian diffusion coefficient, $c$ is the colloidal concentration,
$D_{T}$ is the thermal diffusion coefficient and $T$ is the temperature.
The second term on the right hand side of eq. (\ref{eq:-1}) describes the particle flux induced by a temperature
gradient. From the relation $\vec{J}=c\vec{v}_{T}$, the thermophoretic
velocity can be identified as $\vec{v}_{T}=-D_{T}\nabla T$. In a closed system, a steady-state of the colloidal component is reached when the total flux vanishes:
\begin{equation}
\nabla c=-cS_{T}\nabla T.\label{eq:-2}
\end{equation}

The ratio $S_{T}=D_T/D$, also known as the Soret coefficient, quantifies the strength and direction of colloidal thermophoresis.

In dilute suspensions, colloidal pair-interactions can be neglected and the Stokes-Einstein relation $\gamma D=k_BT$ can be used to relate the diffusion coefficient $D$ to the friction coefficient $\gamma$ of the colloid, where $k_B$ is the Boltzmann constant. As a result, the effective force $\vec{F}_T$ that drives thermophoresis can be written as:

\begin{equation}
\vec{F}_T=\gamma\vec{v}_T=-k_{B}TS_{T}\nabla T.\label{eq:-4}
\end{equation}

Using this expression for $\vec{F}_T$, the steady-state concentration profile of colloids in a temperature gradient can hence be expressed as:

\begin{equation}
\nabla \ln c = \frac{\vec{F}_T}{k_B T}.\label{eq:-5}
\end{equation}
  
Our experimental technique is based on eqs. (\ref{eq:-2}) and (\ref{eq:-5}), which show that the Soret coefficient $S_T$ can be determined from the colloidal concentration profile at steady-state when the temperature gradient is known.

\section{Materials and Methods }

We have performed thermophoretic measurements in aqueous suspensions using three different polystyrene
(PS) particles of varying negative charge, size and surface coating, including Streptavidin
(PS-STV, from microParticles GmbH), Polyethylenglycol-azide groups (PS-PEG-N$_3$, in house)
and Polyethylenglycol-DNA (PS-PEG-DNA, in house\cite{Zupkauskas}). The PS particles
are either dispersed in deionised water (ACROS Organics,
Fisher Scientific), abbreviated as DiW, or custom-made Tris-EDTA buffer ($10\,mM$ Tris-HCl and $1\,mM$ disodium EDTA at pH 8.0, in house), denoted as TE, and diluted down to volume fractions
of about $0.01\%$. Before each experiment, the suspensions
are sonicated for 20 minutes to break up potential aggregates. The hydrodynamic
diameters $d_h$ and zeta potentials $\zeta$ of the PS particles are obtained from Dynamic Light Scattering (DLS) measurements using a Zetasizer (Nano ZS, Malvern).

\begin{figure}
	\centering{}\includegraphics[width=6.5cm,height=7.5cm]{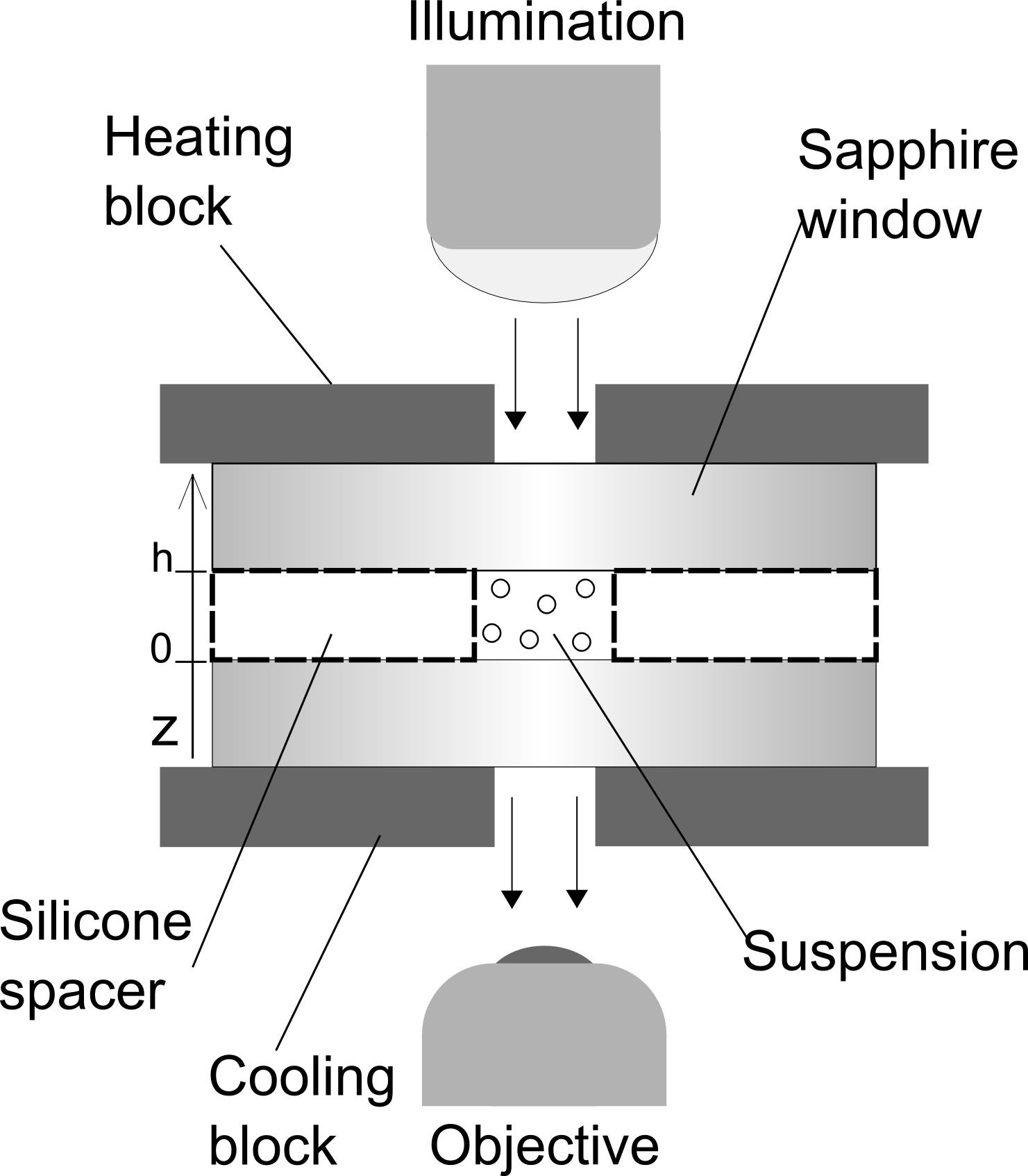}
	
	\caption{Schematic lateral view of the experimental setup (dimensions are not to scale). \label{fig:-1}}
\end{figure}

A schematic diagram
of the setup is shown in fig. \ref{fig:-1}. The cell for the suspension is made of an ultra thin silicone spacer with
a circular hole ($\sim170\,\mu m$ thick, from Silex Silicones LTD),
sandwiched between two sapphire windows (32 x 37 x 0.50mm, from UQG
Optics). Sapphire is optically transparent and a very good heat conductor,
thus guaranteeing a uniform temperature gradient inside
the sample. Upon contact, the silicone
film immediately sticks to the sapphire window due to strong adhesion forces.
A droplet of the suspension ($\sim25\,\mu l$) is then introduced and the second
window is carefully placed on top of the spacer. Moderate pressure is exerted on the top window to squeeze out any excess liquid and to amplify
the adhesion between the windows. The sample
is then transferred to a Nikon Eclipse Ti-E inverted microscope, equipped
with a Ximea MQ013MG-E2 camera with a E2V EV76C560 CMOS sensor. An extra-long working distance objective is used for bright-field imaging, with a numerical aperture of 0.60, corresponding to a depth of focus of about $\sim1.3\,\mu m$. The sample is then mounted onto the microscope
stage and sandwiched between two copper blocks. Both blocks are connected
to PID (proportional integral derivative) controllers and have small central holes for the transmission
of light. To avoid large-scale convection, a uniform temperature
gradient is set up vertically by heating at the top and cooling at
the bottom. For this purpose, the top and bottom blocks are
connected to an electric heater and a water bath; and the corresponding
temperatures are monitored using thermocouples. The time evolution
of the colloidal concentration profile is captured by acquiring images of horizontal slices in 10 minute intervals, spanning the entire
height of the cell. The slices are all equally spaced by a vertical distance of 10 $\mu m$ and averaged over multiple
images to improve the accuracy of the measurement. The local concentration
at each altitude is determined via image analysis using a home-developed MATLAB code, based on a binarisation
method with a high pass filter for contrast and feature size.

\begin{figure}
	\centering{}\includegraphics[width=8cm,height=7.5cm]{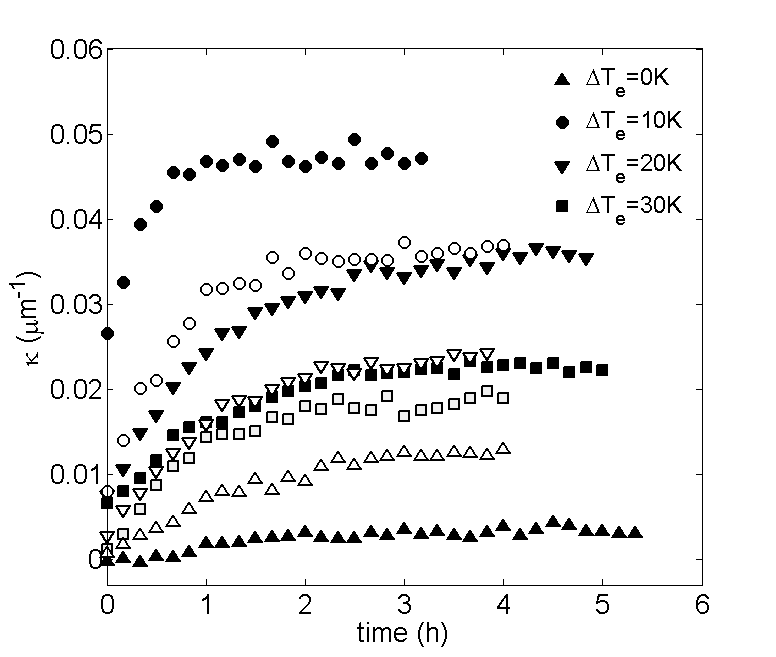}
	
	\caption{Time
		evolution of the decay parameter $\kappa(t)$ for PS-PEG-N$_3$ (full symbols) and PS-STV (empty symbols) in DiW, for varying external temperature differences $\Delta T_e$. The time origin is not absolute, but set to zero at the start of each measurement. \label{fig:-2}}
\end{figure}

\section{Thermophoretic Force Measurement at Steady-State}

In dilute suspensions, a single colloid is subjected to two different forces, the
thermophoretic force $\vec{F}_{T}$ and the
gravitational pull $\vec{F}_{g}=m_{r}\vec{g}$ where $g=9.81\,m\,s^{-2}$.
The reduced mass $m_{r}$ is given by $m_{r}=V_{c}(\rho_{c}-\rho_{w})$,
where $V_{c}$ is the volume of a colloid and $\rho_{c}$ and $\rho_{w}$
are the mass densities of the colloid (PS) and solvent (water), respectively.
Taking into account the gravitational pull of the suspended particle,
the colloidal distribution along the temperature gradient
can hence be written as:

\begin{equation}
\frac{\partial \ln P(z)}{\partial z}=\frac{F{}_{g}}{k{}_{B}T}-S_{T}\frac{\partial T}{\partial z}.\label{eq:-3}
\end{equation}

Here, the colloidal concentration $c(z)$ has been replaced by its corresponding probability distribution $P(z)$. 

In our experiments, we have measured the colloidal concentration profile at steady-state
for different temperature gradients, keeping the bottom block at $20^\circ C$
and raising the temperature of the upper block to a maximum of
$50^\circ C$. Within this narrow temperature range, the expansion of water barely
effects the reduced mass of the colloid and the thermal energy $k_{B}T$
can be assumed constant. In view of eq. (\ref{eq:-3}), the Soret coefficient $S_{T}$ can then be identified as the
negative slope of the curve defined by $\frac{\partial ln\,P}{\partial z}$ vs $\frac{\partial T}{\partial z}$,
allowing a natural elimination of the gravitational pull as a constant offset at $\frac{\partial T}{\partial z}=0$. It is important to note that the colloids must be at steady-state before this measurement technique for $S_{T}$ is applied.
An order of magnitude estimate for the relaxation time required to reach this steady-state is given by
the diffusive time scale $\tau_{D}\sim h^{2}/2D$, where $h$ is the
cell height set by the thickness of the silicone spacer. The diffusion coefficient can be determined
from the Stokes-Einstein relation $D=k_{B}T/\left( 3\pi\eta d_{h}\right)$,
where $d_{h}$ is the hydrodynamic diameter of the particle and $\eta$
is the solvent viscosity, which takes the value of $8.9\,10^{-4}
Pa\,s$ for water at room temperature. The values of $d_h$, $D$ and $\tau_{D}$ are reported in Table \ref{table:-1}.
\\
\begin{table}
\centering{}
\begin{tabular}{|c|c|c|c|}
\hline 
 & PS-STV & PS-PEG-N$_3$ & PS-PEG-DNA \tabularnewline
\hline 
\hline 
$d_{h}\,(nm)$ & 591 & 446 & 491\tabularnewline
\hline 
$D\,(um^{2}s^{-1})$ & 0.84 & 1.11 & 1.01\tabularnewline
\hline 
$\tau_{D}\,(h)$ & 4.78 & 3.62 & 3.97\tabularnewline
\hline  
\end{tabular}
\caption{The values of $D$ are determined from the Stokes-Einstein relation at room temperature, using the values of the hydrodynamic diameters $d_h$.}\label{table:-1}
\end{table}

\begin{figure}
	\centering{}\includegraphics[width=6.5cm,height=11.7cm]{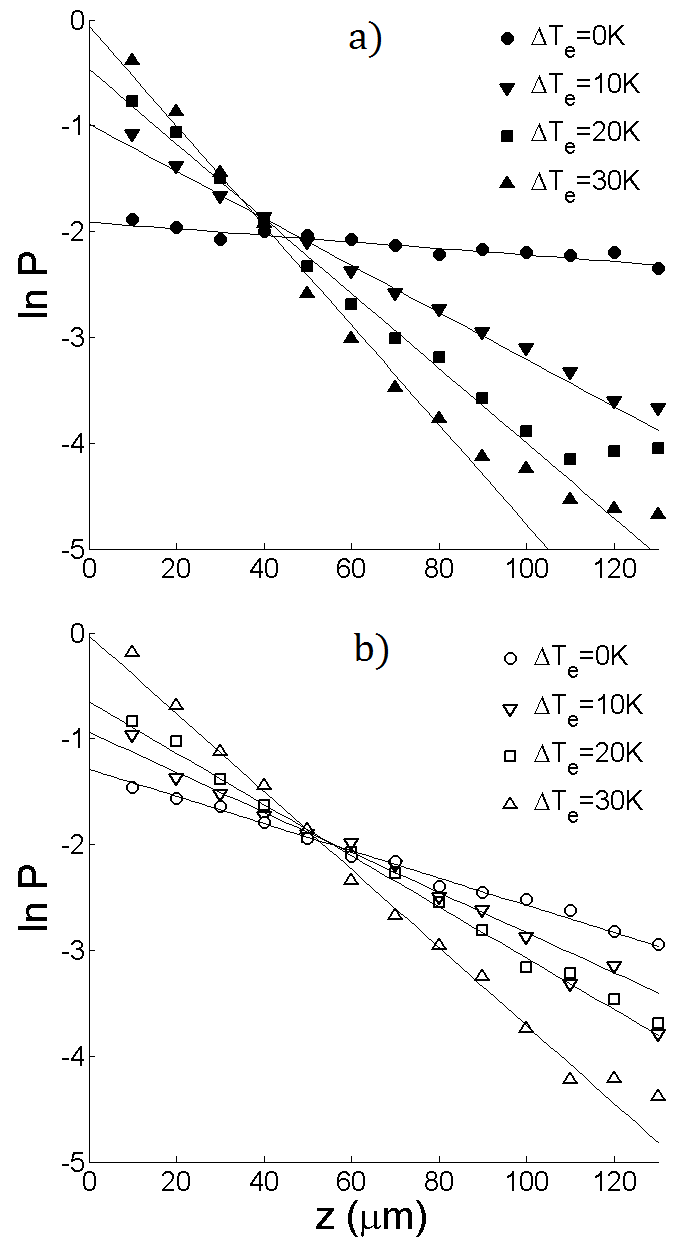}
	
	\caption{Natural logarithm of probability $P(z)$ vs altitude $z$ for varying external temperature differences $\Delta T_e$, for a) PS-PEG-N$_3$ and b) PS-STV, in DiW. The probability $P(z)$ is normalised according to $P(z_i)=c(z_i)/\sum_ic(z_i)$, where $z_i$ are the discrete altitudes inside the bulk range. \label{fig:-3}}
\end{figure}

\begin{figure}
	\centering{}\includegraphics[width=9cm,height=8.7cm]{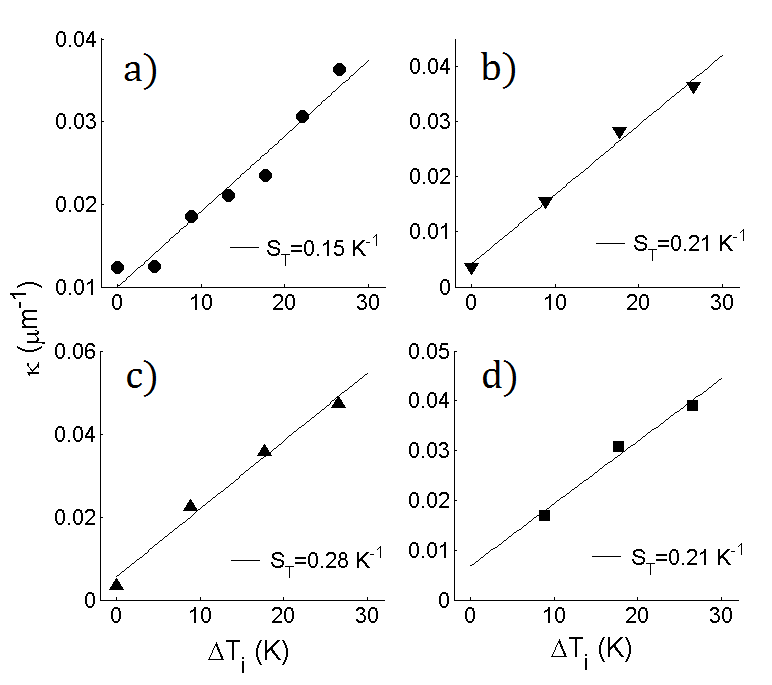}
	
	\caption{Decay parameter $\kappa$ vs internal temperature difference $\Delta T_i$, for a) PS-STV in DiW, b) PS-PEG-DNA in DiW, c) PS-PEG-N$_3$ in DiW and d) PS-PEG-N$_3$ in TE. The values of $\kappa$ are averages over the last 5 measures at steady-state. The corresponding standard errors and propagated errors on $\Delta T_i$ due to an uncertainty in sapphire conductivity are relatively small and therefore not shown. \label{fig:-4}}
\end{figure}

The complete relaxation to steady-state is verified by monitoring
the time evolution of the decay parameter $\kappa(t)$, which we define as the negative slope of the best-fit line to $\ln\,P(z)$ vs $z$. Examples of these time
evolutions are shown in
fig. (\ref{fig:-2}), for PS-PEG-N$_3$ and PS-STV in DiW. For both systems, $\kappa(t)$ reaches a stable value after around 4 hours, which indeed falls into the range of values for $\tau_{D}$ given in Table \ref{table:-1}. 
Fig. (\ref{fig:-3}) shows the plots
of $\ln\,P(z)$ vs $z$ at steady-state for the same systems. It can be seen that the concentration of colloids is highest at the bottom of the cell and falls off rapidly towards the top. The lines
represent best fits to the data over the range of $30$ - $100$ micrometers above the bottom surface. The reason for limiting the fitting to a bulk range is twofold. Apart from colloidal absorption, which tends to be stronger at the bottom due to gravity, lower concentrations give poor statistics close to the top surface. Fluctuations due to poor statistics are visible in fig. \ref{fig:-3} at higher altitudes and become more pronounced with increasing temperature difference. A most notable feature of fig. \ref{fig:-3} is the linearity of $\ln\,P(z)$ in the bulk. In view of eq. (\ref{eq:-5}), this linearity implies that the Soret coefficient $S_T$ is constant throughout the suspension, meaning that the colloids are subjected to a uniform thermophoretic force. At steady-state, the decay parameter $\kappa$ can thus be related to the Soret coefficient via:

\begin{equation}
\kappa=S_T\nabla T+\kappa_g,\label{eq:-21}
\end{equation} 

where $\kappa_g=-F_g/(k_BT)$. Knowing the cell height and the internal temperature difference $\Delta T_i$ inside the suspension, $S_T$ can hence be determined from:

\begin{equation}
S_T=h\frac{\partial\kappa}{\partial\Delta T_i}.\label{eq:-6}
\end{equation} 

The internal temperature difference $\Delta T_i$ differs from the externally applied difference $\Delta T_e$ due to the finite thermal conductivity of sapphire. By treating the sapphire windows and suspension as conducting elements in series, it can be shown that:

\begin{equation}
\Delta T_i=\frac{1}{1+2\frac{\sigma_w h_s}{\sigma_s h}}\Delta T_e,\label{eq:-7}
\end{equation} 

where $h_s$ is the thickness of a sapphire window and $\sigma_s=27.21\,Wm^{-1}K^{-1}$ and $\sigma_w=0.6\,Wm^{-1}K^{-1}$ are the thermal conductivities of sapphire and water, respectively. Using these values, we obtain the relation $\Delta T_i=0.88\,\Delta T_e$. 
\\
\begin{table}
	\centering{}
	\begin{tabular}{|c|c|c|c|}
		\hline 
		in DiW & PS-STV & PS-PEG-N$_3$ & PS-PEG-DNA \tabularnewline
		\hline
		\hline  
		$S_T/d_h\,(K^{-1}\mu m^{-1})$ & 0.25 $\pm$0.02 & 0.63 $\pm$0.04 & 0.43 $\pm$0.03\tabularnewline
		\hline 
		$\zeta\,(mV)$ & -29.5 & -18.9 & -30.6\tabularnewline
		\hline 
	\end{tabular}
	\caption{The errors on $S_T/d_h$ are calculated from the relative sample error of $7\%$.}\label{table:-3}
\end{table}

The plots of $\kappa$ vs $\Delta T_i$ are shown in fig. \ref{fig:-4} for all studied systems. It should be noted that the measurements on PS-STV were performed on different samples, thus explaining the higher noise level in fig. \ref{fig:-4}a. The relative sample error associated with these measurements is $7\%$ and is likely due to a fluctuating pH in DiW. It can be seen that $\kappa(\Delta T_i)$ is approximately linear for each system, indicating that $\vec{F}_T$ is linear in $\nabla T$ and that $S_T$ is rather insensitive to temperature. The values of $S_T$ obtained from eq. (\ref{eq:-6}) are also displayed in fig. \ref{fig:-4} and are exclusively positive, corresponding to a thermophobic behaviour of all studied PS particles. The values of $S_T$ measured in DiW deserve particular attention, as they do not conform with theoretical predictions for charged colloids in aqueous electrolyte solutions\cite{Parola2004,Wurger2010}. More generally, these models predict that the ratio $S_T/d_h$ increases with the magnitude of the zeta potential $\zeta$. For comparison, the measured values of $S_T/d_h$ and $\zeta$ in DiW are given in Table \ref{table:-3}. Although the value of  $\zeta$ may fluctuate in DiW ($\pm 5\,mV$), these measurements show that PS-PEG-N$_3$ clearly has the weakest zeta potential. This is mainly due to the azide ($N_3$) groups on the colloidal surface. Unlike DNA, which carries a net negative charge, the azide groups are neutral and therefore reduce the zeta potential by shifting the hydrodynamic slip plane away from the charged surface. However, the ratio $S_T/d_h$ has been found to be highest for PS-PEG-N$_3$ ($S_T/d_h=0.63\,K^{-1}\mu m^{-1}$), which opposes the theoretical prediction that $S_T/d_h$ increases with the magnitude of $\zeta$. Our measurements thus support the interpretation of thermophoresis as an interfacial effect, but also suggest that surface functionality might lead to a more intricate dependence of $S_T$ on hydrodynamic boundary conditions than predicted by the aforementioned $\zeta$-model.

\section{Colloidal Center of Mass Motion in Response to a Thermophoretic Force}

Although the colloidal steady-state has previously been exploited to determine $S_T$, very little is known about the relaxation process behind this steady-state. The diffusive time scale $\tau_D$ yields a rough estimate for the relaxation time but provides no further insight into the underlying relaxation dynamics. In biological processes however, we are often interested in how a collection of confined particles or molecules relaxes to steady-state under the action of a weak thermodynamic force, the accumulation of biomolecules in out-of-equilibrium pores being an important example \cite{Baaske2007}. A theoretical model is therefore required that allows a more quantitative description of this collective relaxation. Here, we propose the colloidal center of mass (CoM) as a natural candidate for this description, defined as:

\begin{equation}
Z=\frac{\sum_i z_i c(z_i)}{\sum_i c(z_i)}.\label{eq:-10}
\end{equation}

Our theoretical consideration starts from the continuity equation for an effectively one-dimensional, closed system in the absence of particle generation:

\begin{equation}
\frac{\partial P_z}{\partial t}+\frac{\partial j}{\partial z}=0.\label{eq:-8}
\end{equation}

$P_z(z)$ is the linear probability density, satisfying $\int P_z(z) dz=1$. The CoM is then simply related to $P_z(z)$ via $Z = \int z P_z(z) dz$. The corresponding probability flux $j$ is given by:

\begin{equation}
j=-D\frac{\partial P_z}{\partial z}+\frac{F}{\gamma}P_z,\label{eq:-13}
\end{equation}

where $F=F_g+F_T$ is the total force on a colloid. As the temperature variation inside the system is very small ($\Delta T/T\ll 1)$, the diffusion coefficient $D$ and friction coefficient $\gamma$ of a colloid can be taken as approximately constant. With the expression for $j$ given by eq. (\ref{eq:-13}), the continuity equation (\ref{eq:-8}) can then be rewritten in terms of rescaled variables:

\begin{equation}
\frac{\partial P_z}{\partial t'}+\frac{\partial P_z}{\partial z'}-\frac{\partial^2 P_z}{\partial z'^2}=0,\label{eq:-14}
\end{equation}

where $t'=\kappa^2Dt$ and $z'=\kappa z$. Here, we compare our  experimental data to numerical solutions of eq. (\ref{eq:-14}), using a standard PDE-solver (MATLAB). By assuming perfectly reflecting boundaries at $z=0$ and $z=h$, eqs. (\ref{eq:-8}) and (\ref{eq:-13}) can further be used to derive the following equation of motion for the colloidal CoM:

\begin{equation}
F+\Pi-\gamma V_Z=0,\label{eq:-9}
\end{equation}

where $V_Z$ is the CoM velocity. The term $\Pi$ has an entropic nature and is given by:

\begin{equation}
\Pi=-k_BT\left(P_z(h)-P_z(0)\right).  \label{eq:-11}
\end{equation}

As $P_z(z',t')$ only has a simple stationary form at steady-state, there is no straightforward analytical prediction for the time evolution of $\Pi$. 

\begin{figure}
	\centering{}\includegraphics[width=7cm,height=11.6cm]{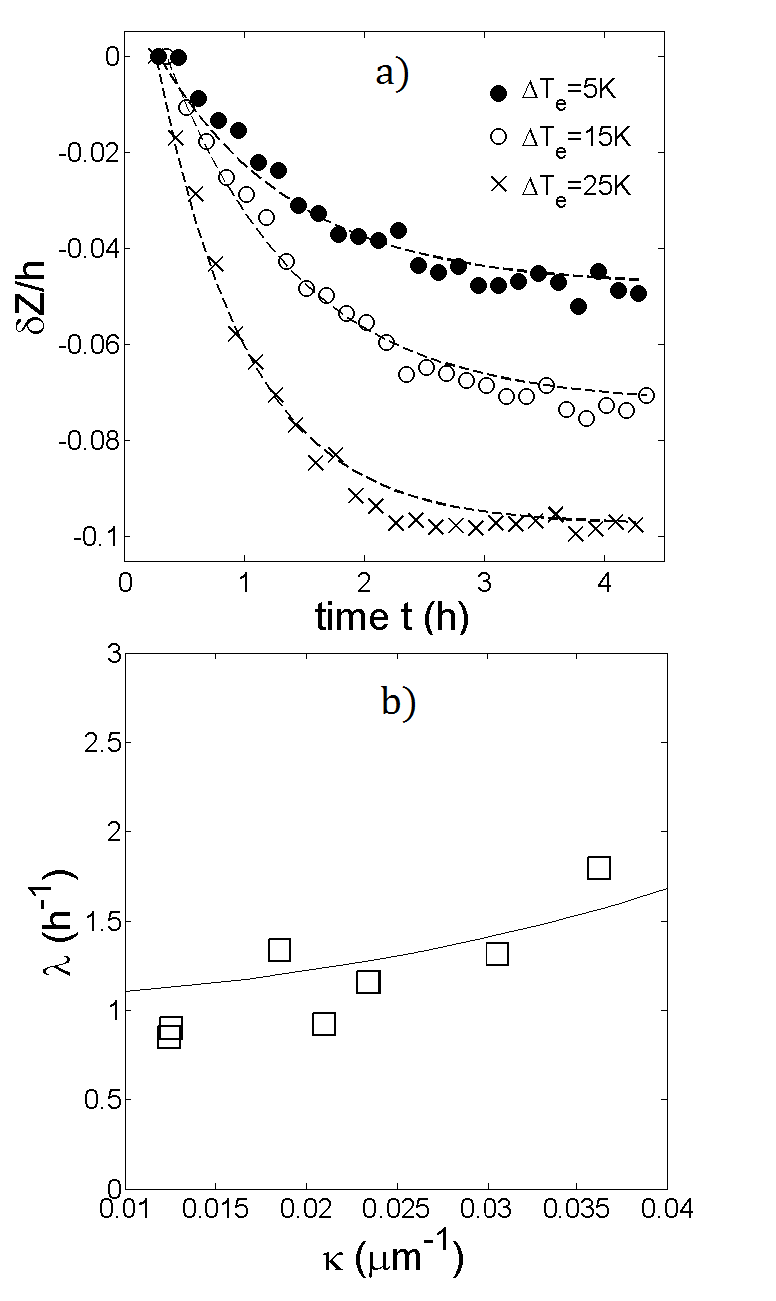}
	
	\caption{a) Observed time evolution of the CoM shift $\delta Z$ of PS-STV in DiW, rescaled with respect to the cell height $h$, for three different values of $\Delta T_e$. The dashed lines are exponential fits from which the values of $\lambda$ are determined. b) Relaxation speed $\lambda$ vs decay parameter $\kappa$ for PS-STV in DiW (squares). The solid line shows the trend of $\lambda(\kappa)$ obtained from numerical solutions of eq. (\ref{eq:-14}), based on reflecting boundaries and using the value of $D$ given for PS-STV in Table \ref{table:-1}. \label{fig:-8}}
\end{figure}

Let us now consider the case where the system is at steady-state. The colloidal CoM has reached a stable position ($V_Z=0$) and the force balance is given by $F+\Pi=0$. The system is then suddenly subjected to a constant perturbation $\delta F$ at time $t=0$, e.g. by increasing the temperature gradient. The resulting CoM shift $\delta Z$ will induce an entropic response $\delta\Pi$ that opposes the external perturbation until a new steady-state is reached. It thus follows that $\delta\Pi$ acts as a restoring force, satisfying $\delta\Pi(\delta Z=0)=0$. To make progress in quantifying the CoM relaxation, we examine the weak perturbation limit by assuming a linear response relation of the form $\delta\Pi\propto\delta Z$. Eq. (\ref{eq:-9}) can then be solved analytically, giving:

\begin{equation}
\delta Z(t)=\delta Z_f\left(1-\exp\left(-\lambda t \right)\right), \label{eq:-12}
\end{equation}

where $\delta Z_f$ is the final CoM shift over the entire system, in response to the perturbation $\delta F$. Following eq. (\ref{eq:-12}), we propose that the temporal decay constant $\lambda$ can be used to quantify the speed of the relaxation to steady-state. In view of eq. (\ref{eq:-9}), $\lambda$ should further satisfy the relation:

\begin{equation}
\gamma\lambda \sim \left| \frac{\delta F}{\delta Z_f}\right| .\label{eq:-20}
\end{equation}

\begin{figure}
	\centering{}\includegraphics[width=8.3cm,height=7.5cm]{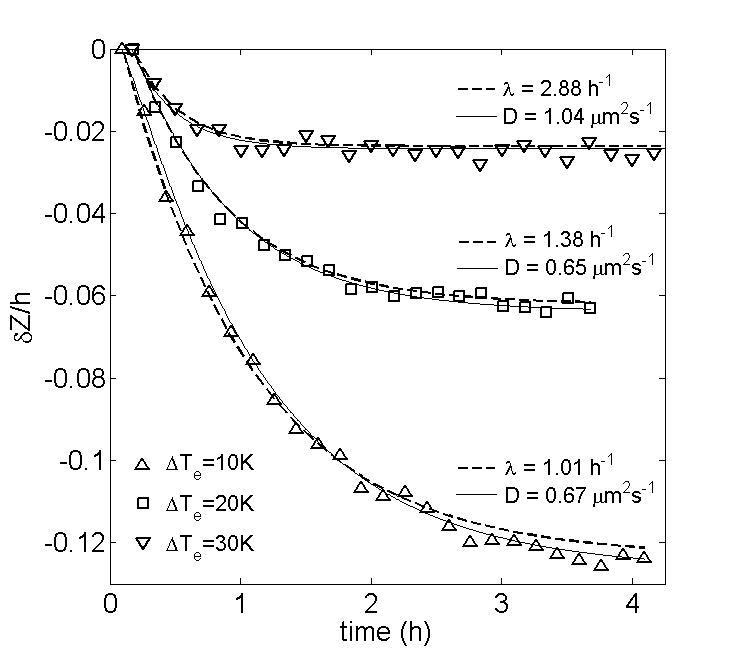}
	
	\caption{Observed time
		evolution of $\delta Z$ for PS-PEG-N$_3$ in TE (symbols), for varying external temperature differences $\Delta T_e$. Full lines correspond to numerical solutions of eq. (\ref{eq:-14}). The legend above each curve shows the value of $\lambda$ obtained from the exponential fit (dashed line) and the optimal value of $D$ used for fitting of the numerical solution. \label{fig:-5}}
\end{figure}

\begin{figure}
	\centering{}\includegraphics[width=7cm,height=6.5cm]{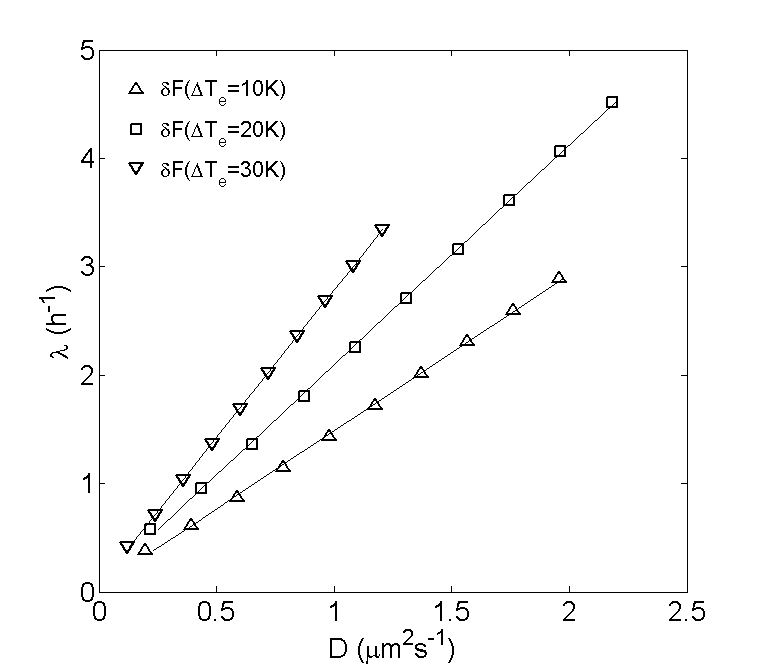}
	
	\caption{Numerical simulation of $\lambda$ vs $D$ (symbols) based on eq. (\ref{eq:-14}), for the given perturbations and initial conditions of PS-PEG-N$_3$ in TE. The full lines are best linear fits.\label{fig:-6}}
\end{figure}

 As eq. (\ref{eq:-12}) relies on reflecting boundaries, it must be noted that some of our colloids suffered from weak surface absorption to the sapphire windows, in particular PS-STV in DiW. Although the number of absorbed colloids saturates during the relaxation, absorption can temporarily perturb the free evolution of the CoM. Furthermore, the system boundaries have to be excluded from the data analysis, leading to deviations from eq. (\ref{eq:-20}) when the CoM is only tracked over a limited bulk range of the system. Nonetheless, we find that the CoM relaxation of PS-STV in DiW is well fitted by an exponential decay. For this system, each relaxation was studied in a separate experiment where an initially uniform distribution of colloids was subjected to a thermophoretic force $F_T$ fixed by the externally applied temperature difference $\Delta T_e$. Three of these relaxations are displayed in fig. \ref{fig:-8}a, together with their exponential fits from which the relaxation speed $\lambda$ is determined. In fig. \ref{fig:-8}b, these values of $\lambda$ are plotted against the corresponding values of $\kappa$, which are directly related to the magnitude of $F_T$ via eq. (\ref{eq:-21}). It can be seen that $\lambda$ tends to increase with $\kappa$, corresponding to shorter relaxation times for stronger thermophoretic forces. The same conclusion is drawn from the trend of $\lambda(\kappa)$ as obtained from numerical solutions of eq. (\ref{eq:-14}). Although assuming reflecting boundaries, the numerical curve (full line) displays a good agreement with the experimental data. The observed trend of $\lambda(\kappa)$ can clearly not be explained by the diffusive time-scale $\tau_D$, which just gives $\lambda=\tau_D^{-1}\propto D$, with no allowance for a dependence on $\kappa$. 

\begin{figure}
	\centering{}\includegraphics[width=8cm,height=7.5cm]{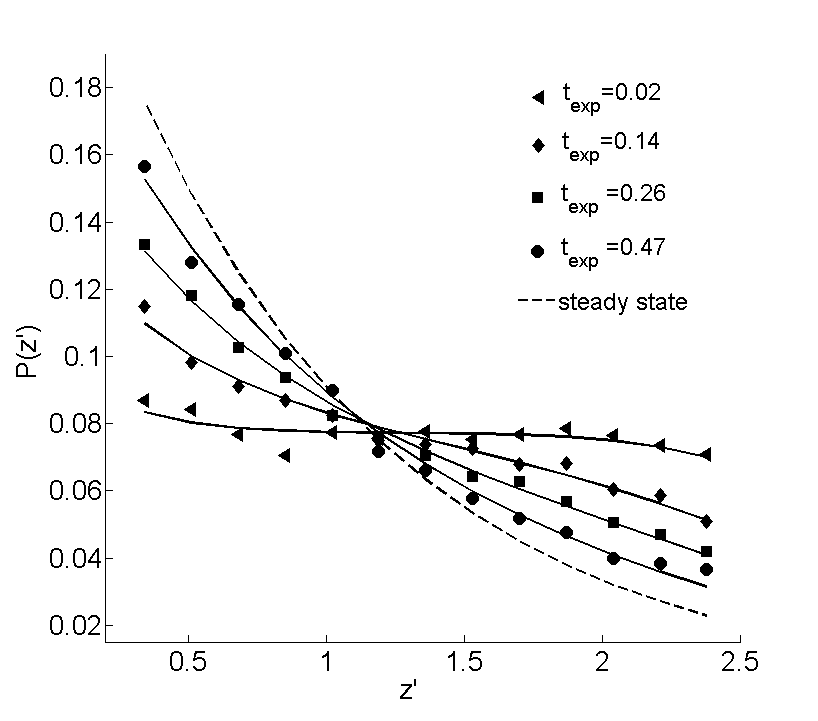}
	
	\caption{a) Probability distribution $P(z')$ vs rescaled altitude $z'$ for PS-PEG-N$_3$ in TE (symbols), at different instants $t_{exp}$ during the relaxation to steady-state for $\Delta T_e=10K$. The full lines are best fits obtained from eq. (\ref{eq:-14}), based on least square fitting. $t_{exp}$ is rescaled w.r.t the total relaxation time. \label{fig:-7}}
\end{figure}

Fig. (\ref{fig:-5}) shows the relaxation of PS-PEG-N$_3$ in TE, for which no surface absorption was observed. The CoM shifts are again very well fitted by our exponential model (dashed lines). This system was measured in a single experiment, so that the initial concentration was uniform for the relaxation at $\Delta T_e=10K$ and subsequently fixed by the previously reached steady-states for the relaxations at $\Delta T_e=20K$ and $\Delta T_e=30K$. For each relaxation, the perturbation $\delta F$ is thus fixed by the difference between the values of $\kappa$ at the final and initial steady-state. In units of $\kappa$, these are given by $\delta F=0.017\,\mu m^{-1}$, $\delta F=0.014\,\mu m^{-1}$ and $\delta F=0.011\,\mu m^{-1}$ in order of increasing $\Delta T_e$. The corresponding values of $\lambda$ are reported in fig. \ref{fig:-5} and indicate that the relaxation speed increased rapidly with the incremental increase of $\Delta T_e$. As the magnitude of $\delta F$ barely changed from one relaxation to the next, the observed increase of $\lambda$ with $\Delta T_e$ in this experiment is mainly related to the initial condition, showing that the CoM relaxes faster when the distance to steady-state $\delta Z_f$ is reduced. 

Due to the absence of absorption, the CoM shift of PS-PEG-N$_3$ can be directly compared to numerical solutions of eq. (\ref{eq:-14}). For this purpose, the solutions (full lines in fig. \ref{fig:-5}) were obtained by using the diffusion coefficient $D$ as a fitting parameter to reproduce the measured value of $\lambda$ most accurately. This is achieved by extracting $D$ from a curve of $\lambda$ vs $D$ based on eq. (\ref{eq:-14}), for each perturbation $\delta F$ and corresponding initial condition. These curves are shown in fig. \ref{fig:-6} and display a linear relationship $\lambda\propto D$ for each relaxation. Interestingly, the thus determined optimal value of $D$ is always smaller than the value of $1.11 \,\mu m^2s^{-1}$ for PS-PEG-N$_3$ obtained from the Einstein relation, the difference being particularly large for the relaxations at $\Delta T_e=10K$ and $\Delta T_e=20K$, where the optimal value is roughly $~0.66 \,\mu m^2s^{-1}$. A possible reason for this discrepancy might be the existence of hydrodynamic effects, giving rise to an additional time-dependent term in eq. (\ref{eq:-13}) that slows down the relaxation and disappears at steady-state. However, fig. \ref{fig:-7} suggests that these effects do not influence the $z'$ -dependence of eq. (\ref{eq:-14}), as the observed shape of the concentration profile $P(z',t')$ of PS-PEG-N$_3$ is always well fitted by a corresponding numerical solution. 

\section{conclusion}
We have introduced a measurement technique for thermophoresis that is based on observing the change of the colloidal steady-state concentration profile with the applied temperature gradient. This method automatically eliminates the gravitational pull and is free of any other external influences, allowing a clean and direct measurement of the Soret coefficient. Our measurements show that the Soret coefficient is rather insensitive to temperature for charged PS particles in aqueous suspensions. Further, the measured thermophroretic force varies linearly with the temperature gradient, supporting the linear-response assumption of the theory of non-equilibrium thermodynamics. Our results also suggest that surface functionality plays an intricate role in thermophoresis that cannot be explained by considerations of surface potential only. We have also investigated the relaxation to steady-state by studying the CoM motion of the colloids. The observed CoM motion is in agreement with a theoretical model that predicts an exponential decay to steady-state. The decay speed $\lambda$ has been found to depend on both the initial condition and the thermophoretic force, with a tendency to increase with the magnitude of the force. This insight cannot be gained from an estimate based on the diffusive time-scale, which only predicts a linear dependence on the diffusion coefficient.   

\section{Acknowledgements}
This work was funded by the Winton Programme for the Physics of Sustainability, Unilever Case and EPSRC (Grant No. 1353070). We are greatly indebted to Emma Talbot and Jurij Kotar for helpful advice and would like to thank Daan Frenkel and Alessio Caciagli for fruitful discussions.

\bibliography{Exp_references}

\begin{thebibliography}{18}%
\makeatletter
\providecommand \@ifxundefined [1]{%
 \@ifx{#1\undefined}
}%
\providecommand \@ifnum [1]{%
 \ifnum #1\expandafter \@firstoftwo
 \else \expandafter \@secondoftwo
 \fi
}%
\providecommand \@ifx [1]{%
 \ifx #1\expandafter \@firstoftwo
 \else \expandafter \@secondoftwo
 \fi
}%
\providecommand \natexlab [1]{#1}%
\providecommand \enquote  [1]{``#1''}%
\providecommand \bibnamefont  [1]{#1}%
\providecommand \bibfnamefont [1]{#1}%
\providecommand \citenamefont [1]{#1}%
\providecommand \href@noop [0]{\@secondoftwo}%
\providecommand \href [0]{\begingroup \@sanitize@url \@href}%
\providecommand \@href[1]{\@@startlink{#1}\@@href}%
\providecommand \@@href[1]{\endgroup#1\@@endlink}%
\providecommand \@sanitize@url [0]{\catcode `\\12\catcode `\$12\catcode
  `\&12\catcode `\#12\catcode `\^12\catcode `\_12\catcode `\%12\relax}%
\providecommand \@@startlink[1]{}%
\providecommand \@@endlink[0]{}%
\providecommand \url  [0]{\begingroup\@sanitize@url \@url }%
\providecommand \@url [1]{\endgroup\@href {#1}{\urlprefix }}%
\providecommand \urlprefix  [0]{URL }%
\providecommand \Eprint [0]{\href }%
\providecommand \doibase [0]{http://dx.doi.org/}%
\providecommand \selectlanguage [0]{\@gobble}%
\providecommand \bibinfo  [0]{\@secondoftwo}%
\providecommand \bibfield  [0]{\@secondoftwo}%
\providecommand \translation [1]{[#1]}%
\providecommand \BibitemOpen [0]{}%
\providecommand \bibitemStop [0]{}%
\providecommand \bibitemNoStop [0]{.\EOS\space}%
\providecommand \EOS [0]{\spacefactor3000\relax}%
\providecommand \BibitemShut  [1]{\csname bibitem#1\endcsname}%
\let\auto@bib@innerbib\@empty
\bibitem [{\citenamefont {Putnam}\ and\ \citenamefont
  {Cahill}(2005)}]{Putnam2005}%
  \BibitemOpen
  \bibfield  {author} {\bibinfo {author} {\bibfnamefont {S.~A.}\ \bibnamefont
  {Putnam}}\ and\ \bibinfo {author} {\bibfnamefont {D.~G.}\ \bibnamefont
  {Cahill}},\ }\bibfield  {title} {\enquote {\bibinfo {title} {{Transport of
  nanoscale latex spheres in a temperature gradient.}}}\ }\href {\doibase
  10.1021/la047056h} {\bibfield  {journal} {\bibinfo  {journal} {Langmuir : the
  ACS journal of surfaces and colloids}\ }\textbf {\bibinfo {volume} {21}},\
  \bibinfo {pages} {5317--23} (\bibinfo {year} {2005})}\BibitemShut {NoStop}%
\bibitem [{\citenamefont {Duhr}\ and\ \citenamefont
  {Braun}(2006{\natexlab{a}})}]{Duhr2006b}%
  \BibitemOpen
  \bibfield  {author} {\bibinfo {author} {\bibfnamefont {S.}~\bibnamefont
  {Duhr}}\ and\ \bibinfo {author} {\bibfnamefont {D.}~\bibnamefont {Braun}},\
  }\bibfield  {title} {\enquote {\bibinfo {title} {{Why molecules move along a
  temperature gradient}},}\ }\href@noop {} {\bibfield  {journal} {\bibinfo
  {journal} {Proceedings of the National Academy of Sciences of the United
  States of America}\ }\textbf {\bibinfo {volume} {2006}} (\bibinfo {year}
  {2006}{\natexlab{a}})}\BibitemShut {NoStop}%
\bibitem [{\citenamefont {Iacopini}, \citenamefont {Rusconi},\ and\
  \citenamefont {Piazza}(2006)}]{Iacopini2006}%
  \BibitemOpen
  \bibfield  {author} {\bibinfo {author} {\bibfnamefont {S.}~\bibnamefont
  {Iacopini}}, \bibinfo {author} {\bibfnamefont {R.}~\bibnamefont {Rusconi}}, \
  and\ \bibinfo {author} {\bibfnamefont {R.}~\bibnamefont {Piazza}},\
  }\bibfield  {title} {\enquote {\bibinfo {title} {{The "macromolecular
  tourist": universal temperature dependence of thermal diffusion in aqueous
  colloidal suspensions.}}}\ }\href {\doibase 10.1140/epje/e2006-00012-9}
  {\bibfield  {journal} {\bibinfo  {journal} {The European physical journal. E,
  Soft matter}\ }\textbf {\bibinfo {volume} {19}},\ \bibinfo {pages} {59--67}
  (\bibinfo {year} {2006})}\BibitemShut {NoStop}%
\bibitem [{\citenamefont {Braibanti}, \citenamefont {Vigolo},\ and\
  \citenamefont {Piazza}(2008)}]{Braibanti2008}%
  \BibitemOpen
  \bibfield  {author} {\bibinfo {author} {\bibfnamefont {M.}~\bibnamefont
  {Braibanti}}, \bibinfo {author} {\bibfnamefont {D.}~\bibnamefont {Vigolo}}, \
  and\ \bibinfo {author} {\bibfnamefont {R.}~\bibnamefont {Piazza}},\
  }\bibfield  {title} {\enquote {\bibinfo {title} {{Does Thermophoretic
  Mobility Depend on Particle Size?}}}\ }\href {\doibase
  10.1103/PhysRevLett.100.108303} {\bibfield  {journal} {\bibinfo  {journal}
  {Physical Review Letters}\ }\textbf {\bibinfo {volume} {100}},\ \bibinfo
  {pages} {108303} (\bibinfo {year} {2008})}\BibitemShut {NoStop}%
\bibitem [{\citenamefont {Duhr}\ and\ \citenamefont
  {Braun}(2006{\natexlab{b}})}]{Duhr2006}%
  \BibitemOpen
  \bibfield  {author} {\bibinfo {author} {\bibfnamefont {S.}~\bibnamefont
  {Duhr}}\ and\ \bibinfo {author} {\bibfnamefont {D.}~\bibnamefont {Braun}},\
  }\bibfield  {title} {\enquote {\bibinfo {title} {{Thermophoretic Depletion
  Follows Boltzmann Distribution}},}\ }\href {\doibase
  10.1103/PhysRevLett.96.168301} {\bibfield  {journal} {\bibinfo  {journal}
  {Physical Review Letters}\ }\textbf {\bibinfo {volume} {96}},\ \bibinfo
  {pages} {168301} (\bibinfo {year} {2006}{\natexlab{b}})}\BibitemShut
  {NoStop}%
\bibitem [{\citenamefont {Dhont}\ \emph {et~al.}(2007)\citenamefont {Dhont},
  \citenamefont {Wiegand}, \citenamefont {Duhr},\ and\ \citenamefont
  {Braun}}]{Dhont2007}%
  \BibitemOpen
  \bibfield  {author} {\bibinfo {author} {\bibfnamefont {J.~K.~G.}\
  \bibnamefont {Dhont}}, \bibinfo {author} {\bibfnamefont {S.}~\bibnamefont
  {Wiegand}}, \bibinfo {author} {\bibfnamefont {S.}~\bibnamefont {Duhr}}, \
  and\ \bibinfo {author} {\bibfnamefont {D.}~\bibnamefont {Braun}},\ }\bibfield
   {title} {\enquote {\bibinfo {title} {{Thermodiffusion of Charged Colloids:
  Single-Particle Diffusion}},}\ }\href@noop {} {\bibfield  {journal} {\bibinfo
   {journal} {Langmuir}\ ,\ \bibinfo {pages} {1674--1683}} (\bibinfo {year}
  {2007})}\BibitemShut {NoStop}%
\bibitem [{\citenamefont {Piazza}\ and\ \citenamefont
  {Guarino}(2002)}]{Piazza2002}%
  \BibitemOpen
  \bibfield  {author} {\bibinfo {author} {\bibfnamefont {R.}~\bibnamefont
  {Piazza}}\ and\ \bibinfo {author} {\bibfnamefont {A.}~\bibnamefont
  {Guarino}},\ }\bibfield  {title} {\enquote {\bibinfo {title} {{Soret Effect
  in Interacting Micellar Solutions}},}\ }\href {\doibase
  10.1103/PhysRevLett.88.208302} {\bibfield  {journal} {\bibinfo  {journal}
  {Physical Review Letters}\ }\textbf {\bibinfo {volume} {88}},\ \bibinfo
  {pages} {208302} (\bibinfo {year} {2002})}\BibitemShut {NoStop}%
\bibitem [{\citenamefont {Piazza}(2003)}]{Piazza2003}%
  \BibitemOpen
  \bibfield  {author} {\bibinfo {author} {\bibfnamefont {R.}~\bibnamefont
  {Piazza}},\ }\bibfield  {title} {\enquote {\bibinfo {title} {{Thermal
  diffusion in ionic micellar solutions}},}\ }\href {\doibase
  10.1080/0141861031000107971} {\bibfield  {journal} {\bibinfo  {journal}
  {Philosophical Magazine}\ }\textbf {\bibinfo {volume} {83}},\ \bibinfo
  {pages} {2067--2085} (\bibinfo {year} {2003})}\BibitemShut {NoStop}%
\bibitem [{\citenamefont {Schimpf}\ and\ \citenamefont
  {Giddings}(1987)}]{Schimpf1987}%
  \BibitemOpen
  \bibfield  {author} {\bibinfo {author} {\bibfnamefont {M.~E.}\ \bibnamefont
  {Schimpf}}\ and\ \bibinfo {author} {\bibfnamefont {J.~C.}\ \bibnamefont
  {Giddings}},\ }\bibfield  {title} {\enquote {\bibinfo {title}
  {{Characterization of thermal diffusion in polymer solutions by thermal
  field-flow fractionation: effects of molecular weight and branching}},}\
  }\href {\doibase 10.1021/ma00173a022} {\bibfield  {journal} {\bibinfo
  {journal} {Macromolecules}\ }\textbf {\bibinfo {volume} {20}},\ \bibinfo
  {pages} {1561--1563} (\bibinfo {year} {1987})}\BibitemShut {NoStop}%
\bibitem [{\citenamefont {Zhang}\ \emph {et~al.}(1999)\citenamefont {Zhang},
  \citenamefont {Briggs}, \citenamefont {Gammon}, \citenamefont {Sengers},\
  and\ \citenamefont {Douglas}}]{Zhang1999}%
  \BibitemOpen
  \bibfield  {author} {\bibinfo {author} {\bibfnamefont {K.~J.}\ \bibnamefont
  {Zhang}}, \bibinfo {author} {\bibfnamefont {M.~E.}\ \bibnamefont {Briggs}},
  \bibinfo {author} {\bibfnamefont {R.~W.}\ \bibnamefont {Gammon}}, \bibinfo
  {author} {\bibfnamefont {J.~V.}\ \bibnamefont {Sengers}}, \ and\ \bibinfo
  {author} {\bibfnamefont {J.~F.}\ \bibnamefont {Douglas}},\ }\bibfield
  {title} {\enquote {\bibinfo {title} {{Thermal and mass diffusion in a
  semidilute good solvent-polymer solution}},}\ }\href {\doibase
  10.1063/1.479498} {\bibfield  {journal} {\bibinfo  {journal} {The Journal of
  Chemical Physics}\ }\textbf {\bibinfo {volume} {111}},\ \bibinfo {pages}
  {2270} (\bibinfo {year} {1999})}\BibitemShut {NoStop}%
\bibitem [{\citenamefont {Duhr}, \citenamefont {Arduini},\ and\ \citenamefont
  {Braun}(2004)}]{Duhr2004}%
  \BibitemOpen
  \bibfield  {author} {\bibinfo {author} {\bibfnamefont {S.}~\bibnamefont
  {Duhr}}, \bibinfo {author} {\bibfnamefont {S.}~\bibnamefont {Arduini}}, \
  and\ \bibinfo {author} {\bibfnamefont {D.}~\bibnamefont {Braun}},\ }\bibfield
   {title} {\enquote {\bibinfo {title} {{Thermophoresis of DNA determined by
  microfluidic fluorescence.}}}\ }\href {\doibase 10.1140/epje/i2004-10073-5}
  {\bibfield  {journal} {\bibinfo  {journal} {The European physical journal. E,
  Soft matter}\ }\textbf {\bibinfo {volume} {15}},\ \bibinfo {pages} {277--86}
  (\bibinfo {year} {2004})}\BibitemShut {NoStop}%
\bibitem [{\citenamefont {Braun}\ and\ \citenamefont
  {Libchaber}(2002)}]{Braun2002}%
  \BibitemOpen
  \bibfield  {author} {\bibinfo {author} {\bibfnamefont {D.}~\bibnamefont
  {Braun}}\ and\ \bibinfo {author} {\bibfnamefont {A.}~\bibnamefont
  {Libchaber}},\ }\bibfield  {title} {\enquote {\bibinfo {title} {{Trapping of
  DNA by Thermophoretic Depletion and Convection}},}\ }\href {\doibase
  10.1103/PhysRevLett.89.188103} {\bibfield  {journal} {\bibinfo  {journal}
  {Physical Review Letters}\ }\textbf {\bibinfo {volume} {89}},\ \bibinfo
  {pages} {188103} (\bibinfo {year} {2002})}\BibitemShut {NoStop}%
\bibitem [{\citenamefont {Jeon}, \citenamefont {Schimpf},\ and\ \citenamefont
  {Nyborg}(1997)}]{Jeon1997}%
  \BibitemOpen
  \bibfield  {author} {\bibinfo {author} {\bibfnamefont {S.~J.}\ \bibnamefont
  {Jeon}}, \bibinfo {author} {\bibfnamefont {M.~E.}\ \bibnamefont {Schimpf}}, \
  and\ \bibinfo {author} {\bibfnamefont {a.}~\bibnamefont {Nyborg}},\
  }\bibfield  {title} {\enquote {\bibinfo {title} {{Compositional effects in
  the retention of colloids by thermal field-flow fractionation.}}}\ }\href
  {\doibase 10.1021/ac9613040} {\bibfield  {journal} {\bibinfo  {journal}
  {Analytical chemistry}\ }\textbf {\bibinfo {volume} {69}},\ \bibinfo {pages}
  {3442--50} (\bibinfo {year} {1997})}\BibitemShut {NoStop}%
\bibitem [{\citenamefont {Piazza}\ and\ \citenamefont
  {Parola}(2008)}]{Piazza2008}%
  \BibitemOpen
  \bibfield  {author} {\bibinfo {author} {\bibfnamefont {R.}~\bibnamefont
  {Piazza}}\ and\ \bibinfo {author} {\bibfnamefont {A.}~\bibnamefont
  {Parola}},\ }\bibfield  {title} {{\selectlanguage {english}\enquote {\bibinfo
  {title} {{Thermophoresis in colloidal suspensions}},}\ }}\href {\doibase
  10.1088/0953-8984/20/15/153102} {\bibfield  {journal} {\bibinfo  {journal}
  {Journal of Physics: Condensed Matter}\ }\textbf {\bibinfo {volume} {20}},\
  \bibinfo {pages} {153102} (\bibinfo {year} {2008})}\BibitemShut {NoStop}%
\bibitem [{\citenamefont {Baaske}\ \emph {et~al.}(2007)\citenamefont {Baaske},
  \citenamefont {Weinert}, \citenamefont {Duhr}, \citenamefont {Lemke},
  \citenamefont {Russell},\ and\ \citenamefont {Braun}}]{Baaske2007}%
  \BibitemOpen
  \bibfield  {author} {\bibinfo {author} {\bibfnamefont {P.}~\bibnamefont
  {Baaske}}, \bibinfo {author} {\bibfnamefont {F.~M.}\ \bibnamefont {Weinert}},
  \bibinfo {author} {\bibfnamefont {S.}~\bibnamefont {Duhr}}, \bibinfo {author}
  {\bibfnamefont {K.~H.}\ \bibnamefont {Lemke}}, \bibinfo {author}
  {\bibfnamefont {M.~J.}\ \bibnamefont {Russell}}, \ and\ \bibinfo {author}
  {\bibfnamefont {D.}~\bibnamefont {Braun}},\ }\bibfield  {title} {\enquote
  {\bibinfo {title} {{Extreme accumulation of nucleotides in simulated
  hydrothermal pore systems.}}}\ }\href {\doibase 10.1073/pnas.0609592104}
  {\bibfield  {journal} {\bibinfo  {journal} {Proceedings of the National
  Academy of Sciences of the United States of America}\ }\textbf {\bibinfo
  {volume} {104}},\ \bibinfo {pages} {9346--51} (\bibinfo {year}
  {2007})}\BibitemShut {NoStop}%
\bibitem [{\citenamefont {W\"{u}rger}(2010)}]{Wurger2010}%
  \BibitemOpen
  \bibfield  {author} {\bibinfo {author} {\bibfnamefont {A.}~\bibnamefont
  {W\"{u}rger}},\ }\bibfield  {title} {\enquote {\bibinfo {title} {{Thermal
  non-equilibrium transport in colloids}},}\ }\href {\doibase
  10.1088/0034-4885/73/12/126601} {\bibfield  {journal} {\bibinfo  {journal}
  {Reports on Progress in Physics}\ }\textbf {\bibinfo {volume} {73}},\
  \bibinfo {pages} {126601} (\bibinfo {year} {2010})}\BibitemShut {NoStop}%
\bibitem [{\citenamefont {Zupkauskas}\ \emph {et~al.}(pted)\citenamefont
  {Zupkauskas}, \citenamefont {Lan}, \citenamefont {Joshi}, \citenamefont
  {Ruff},\ and\ \citenamefont {Eiser}}]{Zupkauskas}%
  \BibitemOpen
  \bibfield  {author} {\bibinfo {author} {\bibfnamefont {M.}~\bibnamefont
  {Zupkauskas}}, \bibinfo {author} {\bibfnamefont {Y.}~\bibnamefont {Lan}},
  \bibinfo {author} {\bibfnamefont {D.}~\bibnamefont {Joshi}}, \bibinfo
  {author} {\bibfnamefont {Z.}~\bibnamefont {Ruff}}, \ and\ \bibinfo {author}
  {\bibfnamefont {E.}~\bibnamefont {Eiser}},\ }\bibfield  {title} {\enquote
  {\bibinfo {title} {{Optically Transparent Dense Colloidal Gels}},}\
  }\href@noop {} {\bibfield  {journal} {\bibinfo  {journal} {Chem. Sci.}\ }
  (\bibinfo {year} {2017, accepted})}\BibitemShut {NoStop}%
\bibitem [{\citenamefont {Parola}\ and\ \citenamefont
  {Piazza}(2004)}]{Parola2004}%
  \BibitemOpen
  \bibfield  {author} {\bibinfo {author} {\bibfnamefont {A.}~\bibnamefont
  {Parola}}\ and\ \bibinfo {author} {\bibfnamefont {R.}~\bibnamefont
  {Piazza}},\ }\bibfield  {title} {\enquote {\bibinfo {title} {{Particle
  thermophoresis in liquids.}}}\ }\href {\doibase 10.1140/epje/i2004-10065-5}
  {\bibfield  {journal} {\bibinfo  {journal} {The European physical journal. E,
  Soft matter}\ }\textbf {\bibinfo {volume} {15}},\ \bibinfo {pages} {255--63}
  (\bibinfo {year} {2004})}\BibitemShut {NoStop}%
\end{thebibliography}%

\end{document}